\documentclass[aip,rsi,reprint,amsmath,amssymb]{revtex4-2}

\usepackage{graphicx}
\usepackage{dcolumn}
\usepackage{bm}
\usepackage{accents}
\usepackage[utf8]{inputenc}
\usepackage[T1]{fontenc}
\usepackage{mathptmx}
\usepackage{etoolbox}
\usepackage{textcomp,mathcomp}
\usepackage[nopatch=footnote]{microtype}
\usepackage{xcolor}

\makeatletter
\def\@email#1#2{%
 \endgroup
 \patchcmd{\titleblock@produce}
  {\frontmatter@RRAPformat}
  {\frontmatter@RRAPformat{\produce@RRAP{*#1\href{mailto:#2}{#2}}}\frontmatter@RRAPformat}
  {}{}
}%
\makeatother
\begin{document}

\preprint{}

\title{Absolute calibration of scintillators for laser-driven ion spectroscopy using \textit{in situ} CR-39 sampling detectors}
\author{Wei Lin$^{1}$, Tzu-Yao Huang$^{1}$, Shih-Fan Yang$^{1}$, Shih-Wei Wang$^{1}$, Kuo-Yu Hsiao$^{1}$, Chun-Han Chen$^{1}$, Yan-Syu Liu$^{1}$, Jun-Yi Chen$^{1}$, Ya-Po Yang$^{1}$, Chen-Kang Huang$^{1,2}$, Hsu-hsin Chu$^{1,2}$, and Chih-Hao Pai$^{1,2}$}
\email{chpai@phy.ncu.edu.tw}
\affiliation{$^1$Department of Physics, National Central University, Taoyuan 320317, Taiwan\\
$^2$Center for High Energy and High Field Physics, National Central University, Taoyuan 320317, Taiwan}%

\date{October 28, 2026}

\begin{abstract}
    Quantitative studies of laser-driven ion acceleration and laser-induced nuclear fusion require accurate measurements of ion yields and energy spectra. Although scintillator screens integrated with Thomson parabola spectrometers provide real-time diagnostics, their nonlinear scintillation response requires absolute calibration. We present an \textit{in situ}, single-shot cross-calibration method using a 2-mm-period CR-39 sampling mask to simultaneously record particle counts and scintillation signals from laser-accelerated proton beams. The method is applied to calibrate a 400-$\mu$m-thick CsI:Tl crystal and a 6-$\mu$m-thick P43 phosphor screen over a broad proton energy range. The measured scintillation responses are compared with a theoretical model combining Birks' law and simulated proton stopping powers. The calibration quantitatively reveals the distinct response characteristics of the two scintillators, including thickness-dependent punch-through effects and constraints on ionization quenching, and demonstrates a practical approach for the absolute calibration of scintillator screens used in laser-driven ion spectroscopy.
\end{abstract}

\maketitle

\section{Introduction}

High-energy ions generated through relativistic laser-plasma interactions have attracted considerable interest for applications including fast ignition and $p$-$^{11}$B fusion, where quantitative characterization of the ion beams is essential. Among the diagnostic techniques used in high-intensity laser experiments, the Thomson parabola spectrometer (TPS) is widely employed to resolve multi-species ion beams and measure their energy spectra~\cite{thomson1907xlvii}. In a TPS, ions pass through a collimator and are subsequently dispersed by static electric and magnetic fields according to their charge-to-mass ratios and velocities. The resulting ion traces are commonly recorded on a scintillating screen. Although scintillating screens provide a convenient and sensitive means of detecting ions over a broad energy range, their nonlinear response to deposited energy requires careful calibration to convert the recorded signal into absolute ion yields and energy spectra.

Several approaches have been developed for the absolute calibration of ion scintillators. Conventional accelerator facilities provide highly stable, monoenergetic ion beams and therefore enable controlled \textit{ex situ} characterization of scintillator response. For example, Schilz et al.~\cite{schilz2024absolute} calibrated a DRZ High scintillating screen using a 150~MeV proton beam, achieving an absolute calibration with an uncertainty of approximately 10\%. Alternatively, scintillators can be cross-calibrated against independent reference detectors, including solid-state nuclear track detectors (SSNTDs), radiochromic films (RCFs), and photostimulable phosphor imaging plates (IPs). Xu \textit{et al.}~\cite{xu2011development}, for example, characterized a BC-430 scintillator by comparing its proton energy spectra with those obtained using CR-39. However, conventional multi-shot cross-calibration methods are susceptible to shot-to-shot fluctuations in laser-driven ion sources, which can introduce additional systematic uncertainty. Spatial sampling techniques have therefore been developed to enable simultaneous measurements with the scintillator and reference detector on a single shot. Tang \textit{et al.}~\cite{tang2020scintillator} used a tungsten resolution grid combined with a partially covered RCF to simultaneously characterize the response of the two detectors, while Cook \textit{et al.}~\cite{cookactive} employed a spatially patterned IP mask in front of a P43 phosphor screen to eliminate the effects of source fluctuations.

Despite these advances, existing single-shot spatial sampling methods remain limited by fundamental constraints. Their quantitative accuracy depends strongly on both the spatial resolution of the sampling structures and the intrinsic properties of the reference detectors. Fabricating fine sampling masks from detectors such as RCFs and IPs is particularly challenging, as cutting these materials can introduce severe edge artifacts resulting from pressure-induced polymerization~\cite{yu2006does} or phosphor flaking~\cite{hammerstrom2006recognition}. Moreover, optical-density-based detectors generally lack discrimination of ion species, which can lead to ambiguity in multi-species ion beams. Their relatively thick protective layers further limit the detection and quantitative characterization of ions over the sub-MeV to multi-MeV energy range.

To address the limitations of existing spatial sampling masks, we demonstrate an \textit{in situ}, single-shot cross-calibration of CsI:Tl and P43 scintillators using a finely patterned, fence-structured CR-39 detector directly within a Thomson parabola spectrometer. Our previously developed automated CR-39 pit-detection algorithm~\cite{lin_cr39}, enhanced by GPU acceleration, enables efficient analysis of high-yield measurements and is well suited to single-shot experiments. The naturally broadband proton spectrum of laser-driven sources allows this approach to extract a continuous, energy-dependent absolute calibration from a single laser shot. The resulting method provides a practical \textit{in situ} diagnostic capability that is complementary to \textit{ex situ} characterizations at conventional accelerators. Correlating the etched-pit statistics in CR-39 with the corresponding scintillation signals recorded by an intensified charge-coupled device (ICCD) camera establishes a direct relationship between the ion fluence measured by CR-39 and the scintillator response. The resulting energy-dependent response is modeled using Birks' law~\cite{birks1951scintillations}, enabling an absolute calibration of the scintillators over a broad multi-MeV proton energy range.

\section{Experimental Setup}

\begin{figure}[htbp]
    \includegraphics[width=\columnwidth]{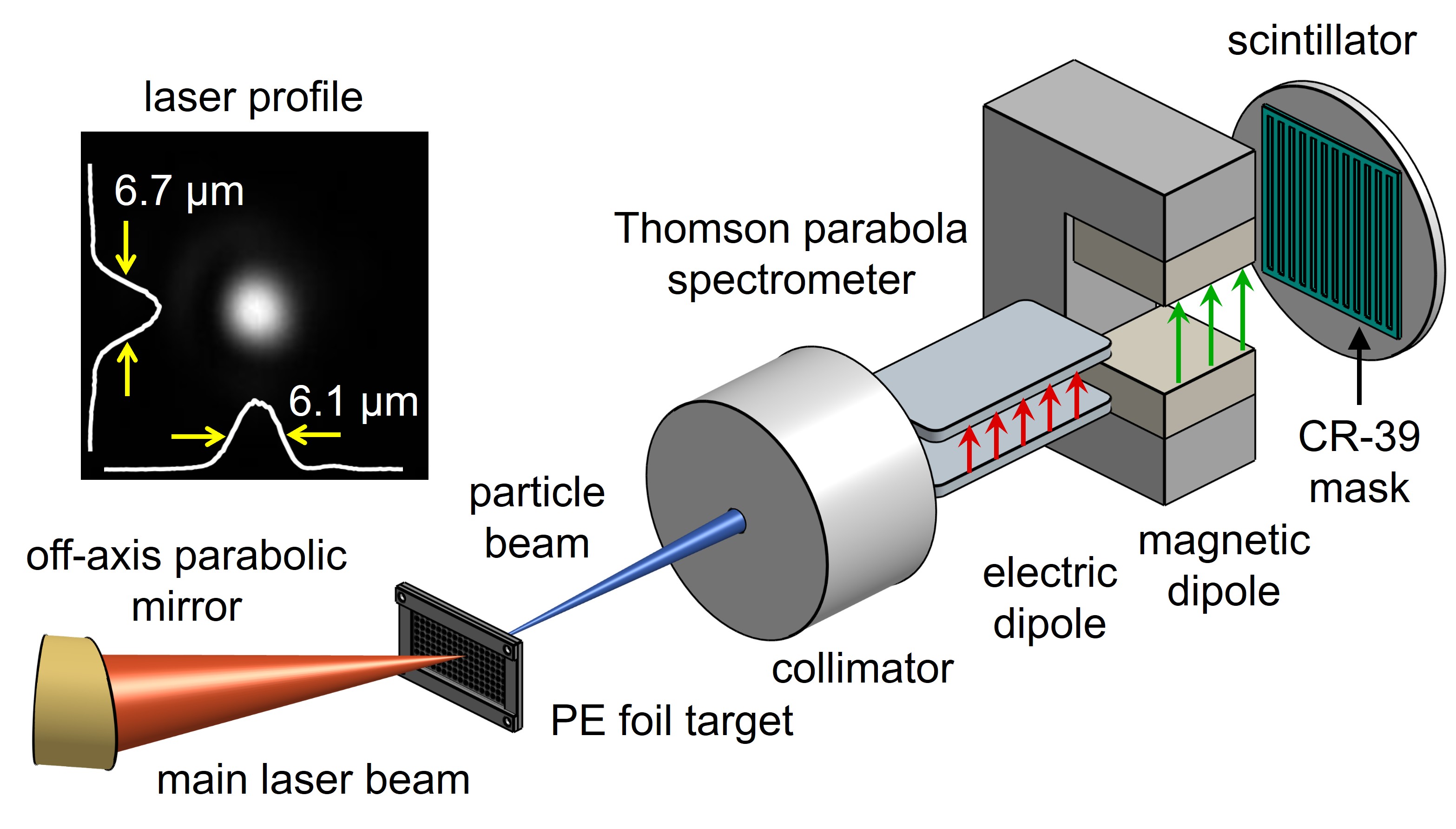}
    \caption{\label{Fig1:setup} Experimental setup for the absolute single-shot cross-calibration of scintillators using a fence-structured CR-39 detector. \emph{Inset:} focal spot image and corresponding intensity profile measured during the experiment. A 2-mm-period CR-39 sampling mask with 1-mm-wide strips and 1-mm-wide openings is placed 5~mm upstream of the scintillator, serving as a spatial sampling mask for the TPS-dispersed ion beam. Red and green arrows indicate the directions of the applied electric and magnetic fields, respectively.}
\end{figure}

The experiment was conducted using the 100-TW Ti:sapphire laser system at National Central University~\cite{Hung2014APB}, as schematically illustrated in Fig.~\ref{Fig1:setup}. An 800-nm, $p$-polarized laser pulse with a full width at half maximum (FWHM) duration of 43~fs and a temporal contrast ratio of $10^{-9}$ at $-250$~ps was focused onto the target by an $f/6$ off-axis parabolic mirror (OAP). The focal spot had a geometric-mean FWHM diameter of 6.4~$\mu$m, with 81.6\% of the measured laser energy contained within the Gaussian-fit profile, resulting in a peak intensity of $5.0\times10^{19}$~W/cm$^2$. An 11-$\mu$m-thick polyethylene (PE) foil was used as the target and positioned at the focal plane. The foil was mounted on a holder designed to withstand approximately 150 laser shots.

Ion beams generated via the target normal sheath acceleration (TNSA) mechanism were characterized using a downstream TPS aligned along the target normal. The customized TPS consists of a collimator located 30~cm from the target, defining a collection divergence angle of $\sim$1~mrad, followed by electric and magnetic dipoles with field strengths of 200~kV/m and 0.31~T, respectively. The active detection area includes the zero-deflection point, providing a sufficiently broad field of view to resolve the continuous ion energy spectrum from 0.3 to 7~MeV.

The scintillators investigated in this study were a P43 phosphor screen and a CsI:Tl scintillator. The P43 screen consists of a 6-$\mu$m-thick terbium-doped gadolinium oxysulfide (Gd$_2$O$_2$S:Tb) phosphor layer with a peak emission wavelength of 545~nm, whereas the CsI:Tl scintillator is a 400-$\mu$m-thick single crystal of thallium-doped cesium iodide with a peak emission wavelength of 550~nm. The scintillation light from both materials was recorded by an ICCD camera with spectral sensitivity covering their respective emission bands.

For the absolute single-shot cross-calibration, an 800-$\mu$m-thick CR-39 track detector was laser-machined into a periodic fence structure with 1-mm-wide CR-39 strips and 1-mm spacing between adjacent strips. The fence-structured CR-39 was positioned 5~mm upstream of the scintillator. For the $\sim$1-mrad collection divergence, this separation corresponds to a lateral broadening of only $\sim 5~\mu\mathrm{m}$, negligible relative to the 1-mm strip and opening widths. Following irradiation, the CR-39 samples were chemically etched in a 7.25~M NaOH solution at $(80\pm1)$~\textdegree C under continuous stirring to ensure uniform pit development. Optical micrographs of the etched CR-39 samples were subsequently acquired using a TASLImage$^\mathrm{TM}$ microscope system~\cite{taslimage} and processed using our automated pit-detection algorithm~\cite{lin_cr39}.

\section{Data Analysis and Calibration}

\begin{figure}[htbp]
    \includegraphics[width=\columnwidth]{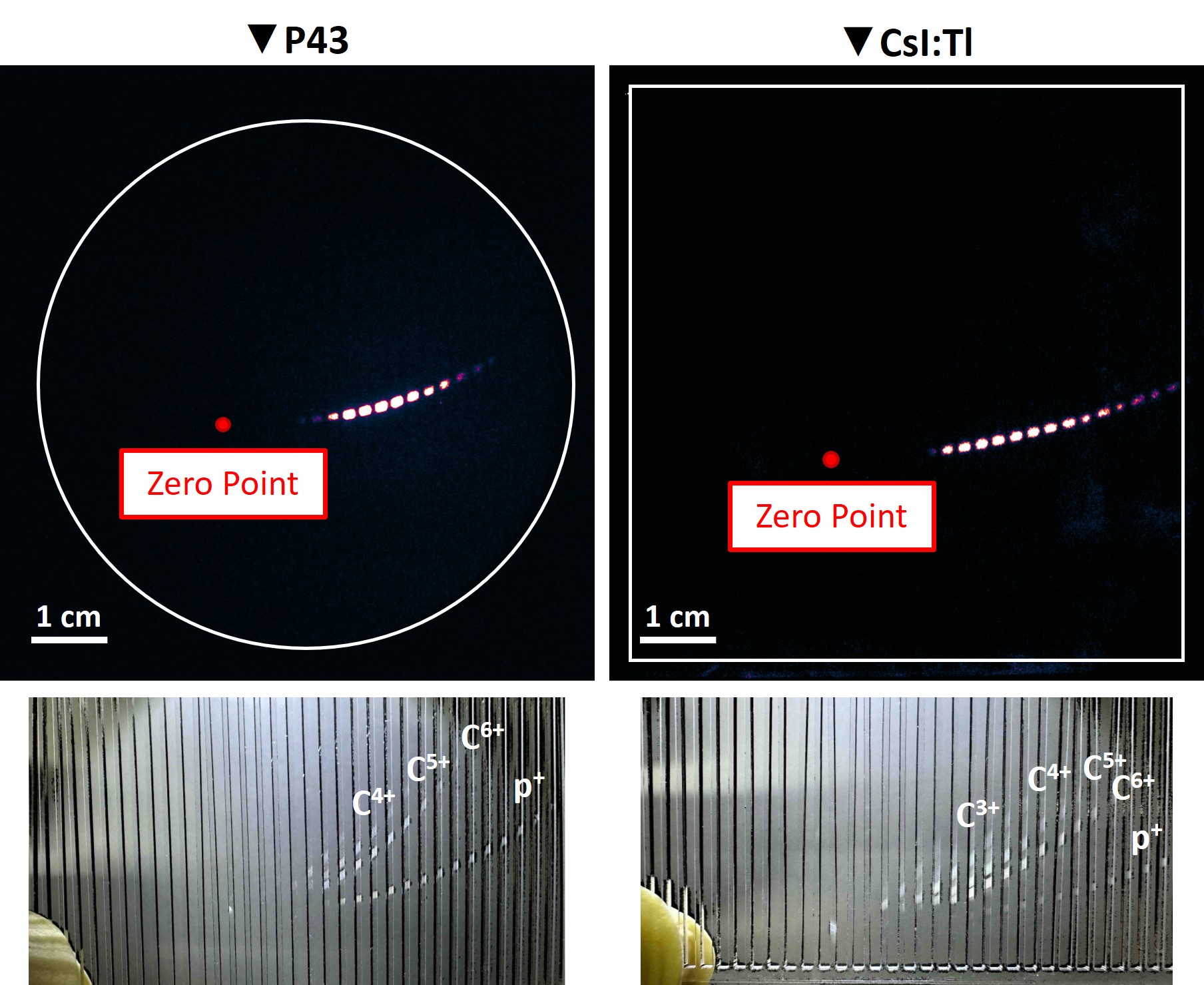}
    \caption{\label{Fig:calibration} Spatially correlated proton signals from the fence-structured CR-39 detector (bottom row) and the scintillator (top row). The zero-deflection points are indicated by red dots. Protons are dispersed by the TPS and their resulting segmented parabolic traces are captured by the ICCD camera. Although heavier species such as carbon ions are observed in the etched CR-39 samples, the absolute calibration focuses exclusively on the dominant proton signals.}
\end{figure}

The calibration experiments were performed independently for the P43 and CsI:Tl scintillators. For each scintillator, the CR-39 mask and the corresponding scintillator were measured simultaneously in a single laser shot. As illustrated in Fig.~\ref{Fig:calibration}, a fraction of the multi-MeV protons is stopped by the CR-39 strips and recorded as tracks, while the remaining protons pass through the slots between strips and reach the downstream scintillator, producing segmented parabolic traces on the ICCD camera. This spatial sampling configuration establishes a direct correspondence between the absolute proton fluence recorded by CR-39 and the scintillation signal generated by the transmitted protons. The absolute number of protons intercepted by the CR-39 strips was determined from the etched-pit statistics using our automated algorithm~\cite{lin_cr39}, which employs directional gradient matching and peak separation approaches to accurately resolve overlapping pits and decouple surface defects. The corresponding scintillation signals were quantified from the ICCD pixel counts, as described in the following sections.

\subsection{Conversion of ICCD Counts to Photon Yield}

The scintillation light was recorded using a PI-MAX4 ICCD camera (Princeton Instruments, USA) with a 1-ms gate duration chosen to capture the majority of the scintillator emission while minimizing background noise from other light sources. To correlate the incident proton fluence with the scintillation signal, the integrated ICCD pixel count, $C$, was converted to the total photon yield, $Y$, assuming isotropic scintillation emission:
\begin{equation}\label{eq:iccd_conversion}
    C = Y \cdot \eta_{\mathrm{geo}} \cdot T_{\mathrm{opt}} \cdot K_{532,G=1} \cdot \mathcal{R}_{\mathrm{spec}} \cdot G.
\end{equation}
Here, $\eta_{\mathrm{geo}}=\Omega/4\pi$ is the fraction of the total solid angle ($\Omega$) subtended by the collection optics, corresponding to $\eta_{\mathrm{geo}}=(1.099\pm0.018)\times10^{-4}$ in our imaging system. A 532-nm continuous-wave (CW) laser was used to calibrate both the optical path transmission and the ICCD photon-to-count conversion factor at $G=1$. The measured optical transmission was $T_{\mathrm{opt}}=(77.9\pm3.4)\%$ at 532~nm and was assumed to be constant over the scintillator emission bands. $K_{532,G=1}$ is the ICCD conversion factor, with a value of $(0.682\pm0.060)$~counts/photon. The dimensionless factor $\mathcal{R}_{\mathrm{spec}}$ accounts for the spectral correction and is calculated by weighting the product of the filter transmission and relative ICCD spectral response by each scintillator's emission spectrum. Normalization to the response at 532~nm yields values of 0.7079 for CsI:Tl and 0.07695 for P43. $G$ denotes the calibrated relative ICCD intensifier gain, normalized to unity at the reference setting.

These calibrations yield conversion factors of $(2.10\pm0.45)\times10^4$~photons/count for the P43 screen recorded at $G=10.61\pm2.18$ and $(2.42\pm0.24)\times10^4$~photons/count for the CsI:Tl crystal recorded at $G=1$. The stated uncertainties were obtained by propagating the relevant calibration uncertainties through Eq.~(\ref{eq:iccd_conversion}).

\subsection{Scintillator Calibration and Fitting}

\begin{figure}[htbp]
    \includegraphics[width=0.9\columnwidth]{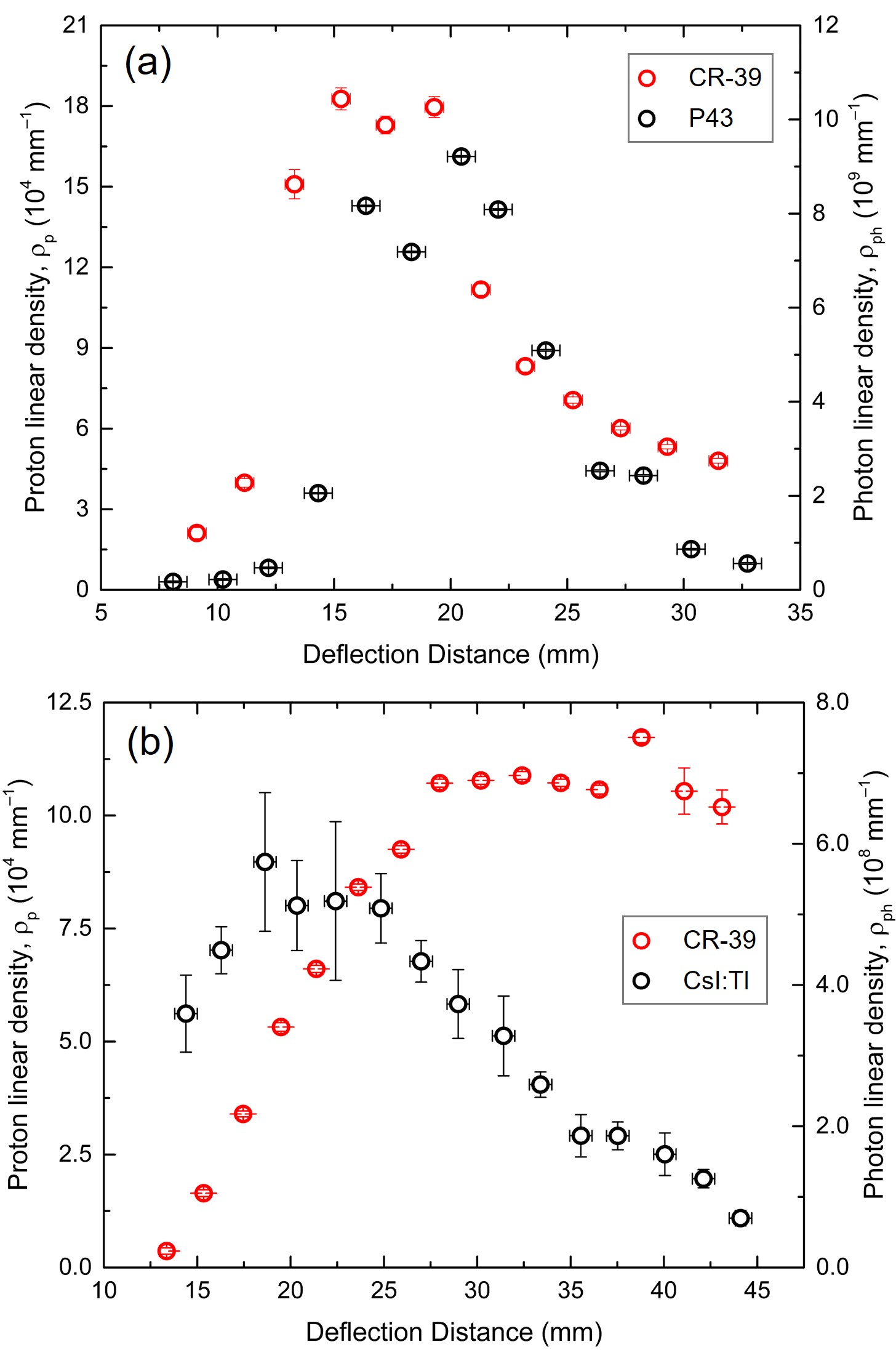}
    \caption{\label{Fig:data} Resolved proton signal distributions along the Thomson parabola traces. The proton linear density (red circles, in $\mathrm{protons/mm}$) extracted from the CR-39 strips and the corresponding photon linear density (black circles, in $\mathrm{photons/mm}$) from the adjacent scintillator openings are plotted against the magnetic deflection distance for (a) the P43 phosphor screen and (b) the CsI:Tl crystal. Horizontal error bars indicate the deflection interval spanned by each 1-mm spatial-integration column, while vertical error bars represent the uncertainties in the column-wise signal extraction.}
\end{figure}

Both the photon signals from the scintillators and the etched pit counts from the CR-39 mask are integrated column-wise at the center of each parabolic segment. The absolute number of protons $N_p$ intercepted by a CR-39 strip of width $w_{\mathrm{strip}} = 1$~mm provides a local measurement of the proton linear density, $\rho_p = N_p / w_{\mathrm{strip}}$. The resulting linear densities are shown in Fig.~\ref{Fig:data}. As the proton fluence along the TPS trace varies smoothly over the 2-mm period of the mask, linear interpolation of the CR-39-derived proton linear densities is used to estimate $\rho_p$ at the spatial coordinates of the corresponding scintillator openings. Based on the magnetic deflection relations of the Thomson parabola, referenced to the zero-deflection points shown in Fig.~\ref{Fig:calibration}, the spatial coordinates are converted into the corresponding incident proton energies. The absolute scintillation response, $S(E)$, defined as the total scintillation photon yield per incident proton, is then obtained as the ratio of the photon linear density $\rho_{\text{ph}}$ to the interpolated proton linear density $\rho_p$:
\begin{equation} \label{eq:response}
    S(E) = \frac{\rho_{\text{ph}}(E)}{\rho_p(E)} = \int_0^{x_0} \frac{A \, (dE/dx)}{1 + kB \, (dE/dx)} \, dx,
\end{equation}
where $x_0$ represents the physical thickness of the scintillator layer, $A$ is the scintillation efficiency, and $kB$ is the Birks constant~\cite{birks1951scintillations}. The proton stopping power, $dE/dx$, and its variation with penetration depth are simulated using the Geant4 toolkit (version 11.3.0)~\cite{agostinelli2003geant4}. The simulation models the scintillators at their nominal densities ($7.32~\text{g/cm}^3$ for P43 and $4.51~\text{g/cm}^3$ for CsI:Tl) using the standard electromagnetic physics list (\texttt{G4EmStandardPhysics}) alongside ion and decay physics modules. Secondary particles are explicitly tracked with a production threshold (range cut) of $1~\mu\text{m}$. The integral in Eq.~(\ref{eq:response}) therefore accounts for the energy deposited throughout the scintillator, including the transition from complete energy deposition to partial energy deposition when protons begin to pass through the scintillator layer.

\begin{figure}[htbp]
    \includegraphics[width=0.9\columnwidth]{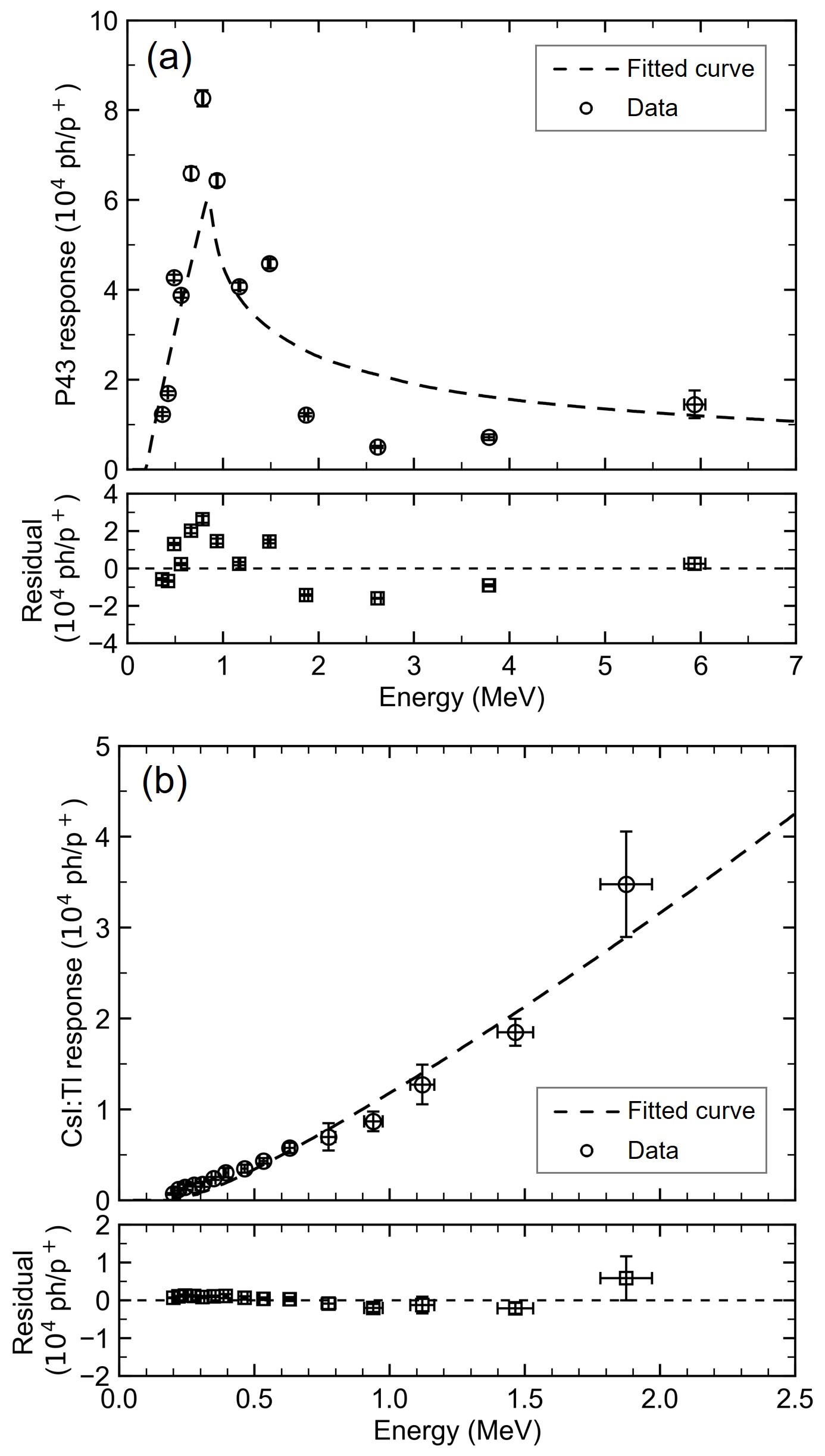}
    \caption{\label{Fig:response} Absolute scintillation response as a function of proton energy for (a) the P43 phosphor screen and (b) the CsI:Tl crystal. The dashed curves represent the theoretical fits based on Birks' law coupled with energy deposition profiles. The lower panels show residuals, defined as the experimental response minus the fitted response at each measured energy. The residual error bars are propagated from the plotted experimental uncertainties.}
\end{figure}

As shown in Fig.~\ref{Fig:response}, fitting Eq.~(\ref{eq:response}) to the experimental response yields the scintillation parameters. For the CsI:Tl crystal, the fitted scintillation efficiency is $A = (3.71 \pm 2.88) \times 10^4$~photons/MeV and the Birks constant is $kB = (2.90 \pm 3.66) \times 10^{-3}$~mm/MeV. For the P43 screen, the fit yields $A = (8.08 \pm 2.48) \times 10^4$~photons/MeV and a Birks constant consistent with zero, $kB = (0.00 \pm 2.35) \times 10^{-4}$~mm/MeV, with the resulting response showing an approximately linear dependence on the incident energy below the punch-through threshold. These fit uncertainties and the error bars shown in Figs.~\ref{Fig:data} and~\ref{Fig:response} do not include the conversion uncertainty discussed in Sec.~III~A.

The theoretical model reproduces the principal energy-dependent trends of both scintillators. The CsI:Tl response exhibits a monotonic, approximately linear increase with incident proton energy, consistent with protons being fully stopped within the relatively thick crystal. In contrast, the P43 response exhibits a pronounced maximum near $0.8$~MeV, corresponding approximately to the punch-through threshold, above which an increasing fraction of the incident proton energy is deposited beyond the active phosphor layer. However, substantial deviations from the model remain, particularly for P43.

For the thin P43 phosphor layer, the fitted Birks constant is consistent with zero within the experimental uncertainty, indicating that no measurable quenching effect is required to describe the response over the range of stopping powers encountered by MeV-class protons. Consequently, below the punch-through threshold near 0.8~MeV (Fig.~\ref{Fig:response}(a)), the scintillation yield of P43 is approximately proportional to the deposited energy. The experimental deviations from the fitted response, however, are substantially larger than the statistical uncertainties of the individual data points. This indicates the presence of systematic variations that are not captured by the idealized model. In particular, P43 screens fabricated by sedimentation consist of a granular layer of phosphor microcrystals, for which local variations in coating density, layer thickness, surface morphology, and optical scattering can produce spatially nonuniform scintillation signals~\cite{kilcoyne2003interferometer}. These effects may contribute to the observed deviations from the calculated response.

In contrast, the single-crystal CsI:Tl detector does not exhibit the granular structures associated with the P43 phosphor layer. Its relatively large thickness also ensures that the incident protons are fully stopped over the investigated energy range, resulting in a monotonic response governed primarily by the total energy deposited in the crystal. Despite this physical uniformity, the fitted CsI:Tl parameters have substantially larger relative uncertainties than those obtained for P43. Because the CsI:Tl response exhibits only weak nonlinearity for protons below 2~MeV, the data provide limited sensitivity to the Birks constant. Consequently, $kB$ is poorly constrained and strongly correlated with the fitted scintillation efficiency $A$, producing the relatively large uncertainties of both parameters.

Nevertheless, the spatial sampling provided by the 2-mm-period CR-39 mask enables the response profiles to be resolved with sufficient detail to identify the transition from the fully stopped to punch-through regime of the P43 screen. The single-shot experimental configuration enables each scintillator to be calibrated against the simultaneously sampled proton fluence without relying on shot-to-shot reproducibility. Although the systematic deviations observed for P43 limit the accuracy of a simple response model based solely on the idealized energy-deposition profile, the CsI:Tl crystal exhibits a smoother response with smaller deviations from the fitted model and therefore shows greater potential for quantitative ion spectrometry over the investigated energy range.

Overall, this calibration experiment establishes an \textit{in situ} method for determining the absolute response of scintillating detectors used in Thomson parabola spectrometers. The results also suggest several directions for further improving the accuracy and applicability of the response model. Measurements over a broader range of proton energies and stopping powers would enable a more robust determination of $kB$ for CsI:Tl. Characterization of the surface morphology and thickness uniformity of phosphor screens could further quantify and correct systematic variations associated with granular scattering and coating nonuniformity. Finally, extending the present cross-calibration approach to heavier ion species would enable characterization of scintillation response across a broader range of particle species and stopping powers.

\section{Conclusion}

We have demonstrated an absolute, single-shot cross-calibration method for ion scintillators using a 2-mm-period CR-39 mask. By simultaneously sampling the laser-driven proton spectrum with CR-39 and scintillators, the method eliminates the need to compare measurements from separate shots and substantially reduces systematic uncertainties associated with shot-to-shot fluctuations. The continuous proton spectrum, extending to energies of $\sim$7~MeV, enables the scintillation response to be characterized over a broad energy range within a single laser shot.

The absolute responses of a 400-$\mu$m-thick CsI:Tl crystal and a 6-$\mu$m-thick P43 phosphor screen were quantitatively determined. The measured responses were modeled by combining Birks' law with proton energy-deposition profiles calculated using Geant4. The P43 response was found to be approximately proportional to the deposited energy below its punch-through threshold, with no measurable quenching over the investigated energy range, whereas the CsI:Tl crystal exhibited a monotonic, approximately linear response for MeV-class protons. The P43 data exhibited larger deviations from the idealized response model, whereas the CsI:Tl crystal showed a smoother response over the measured range.

This \textit{in situ} cross-calibration approach provides a practical means of establishing the absolute scintillation response under the same experimental conditions as laser-driven ion measurements. It therefore improves the quantitative capability of Thomson parabola spectrometers and provides a general framework for calibrating scintillator-based ion diagnostics in laser-plasma experiments.

\section{Acknowledgements}
This work was supported by National Science and Technology Council in Taiwan (Grant No. NSTC 115-2119-M-008-007 and 115-2112-M-008-015).

\bibliographystyle{aipnum4-2}
\bibliography{ref}

\end{document}